\documentclass[preprint,12pt]{elsarticle}

\usepackage{amssymb}
\usepackage{amsmath}

\usepackage[hidelinks]{hyperref}

\journal{Nuclear Instruments and Methods in Physics Research Section A: Accelerators, Spectrometers, Detectors and Associated Equipment}

\begin{document}

\begin{frontmatter}



\title{Optimization and Underground Implementation of the KAPAE Phase II Detector for an Invisible New Particle Search in Positronium Decay}

\author[label1]{Dongwoo Jeong}

\affiliation[label1]{organization={The Center for High Energy Physics, Kyungpook National University},
             city={Daegu},
             postcode={41566},
             country={Korea}}

\affiliation[label2]{organization={Department of Physics, Kyungpook National University},
             city={Daegu},
             postcode={41566},
             country={Korea}}

\affiliation[label3]{organization={Center for Underground Physics, Institute for Basic Science},
             city={Daejeon},
             postcode={34126},
             country={Korea}}

\affiliation[label4]{organization={Advanced Radiation Technology Insititute, Korea Atomic Energy Research Insititute},
             city={Jeongup-Si},
             postcode={56212},
             country={Korea}}
             
\affiliation[label5]{organization={Central Research Institute, Korea Hydro \& Nuclear Power Co., Ltd.},
             city={Daejeon},
             postcode={34101},
             country={Korea}}

\affiliation[label6]{organization={State Key Laboratory of Functional Crystals and Devices, Shanghai Institute of Ceramics, Chinese Academy of Sciences},
             city={Shanghai},
             postcode={201899},
             country={China}}

\author[label1,label4]{Jaeyoung Cho}
\author[label2]{Doohyeok Lee}
\author[label3]{Jaehyeok Kim}
\author[label5]{HyeoungWoo Park}
\author[label6]{Yun-Tao Wu}

\author[label1,label2]{H. J. Kim\corref{cor1}}
\ead{hongjoo@knu.ac.kr}
\cortext[cor1]{Corresponding author}

\begin{abstract}
The KAPAE Phase II detector was developed to search for invisible decays of positronium as a probe of physics beyond the Standard Model. The detector geometry was optimized with Geant4 simulations, and the scintillation and readout performance were evaluated experimentally. The temperature dependence of the BGO scintillator and several readout configurations were studied to find the operating condition giving the best energy resolution. The background level was then measured both at the surface and underground. At the surface, the positron-gamma coincidence background was of order 1~Hz, with occasional bursts up to about 65~Hz. Underground, it decreased to below about $3 \times 10^{-4}$~Hz, more than three orders of magnitude lower.
\end{abstract}


\end{frontmatter}



\section{Introduction}
\label{sec:Intro}

Positronium (Ps) is a bound state of an electron and a positron. It provides a simple leptonic system for precision tests of quantum electrodynamics~\cite{karshenboim} and for searches for rare decay channels beyond the Standard Model~\cite{rubbia2004,gninenko2002}. Positronium is purely leptonic, and the photon multiplicities of its two known annihilation modes are precisely predicted: two photons for para-positronium (p-Ps) and three photons for ortho-positronium (o-Ps). Missing or anomalous energy in the decay is therefore a possible signature of an undetected particle. Depending on which annihilation mode is affected, an invisible new particle search in positronium decay probes different classes of hidden-sector states. A partial invisible decay of p-Ps mainly probes a light dark photon~\cite{darkp2}. A total invisible decay of o-Ps can probe axion-like particles~\cite{oPsALP}, milli-charged particles~\cite{MCP32}, mirror-sector particles~\cite{mirror1986}, and other light exotic states.

The Kyungpook National University Advanced Positronium Annihilation Experiment (KAPAE) was set up to pursue this search with a compact scintillation detector. KAPAE Phase I used a highly segmented BGO calorimeter with dual-end SiPM readout, suitable for measuring visible multi-photon decay topologies with reasonable angular information~\cite{KAPAE1jinst,KAPAE1op}. However, the large number of channels and the inactive regions between segments are not well suited to an invisible new particle search, which instead calls for high photon containment and a small uninstrumented area.

Invisible new particle searches in positronium decay have also been carried out with other techniques, including the J-PET and cylindrical NaI-array experiments~\cite{jpet,jpetfeas,apex} and HPGe-based measurements~\cite{hpge}. Closest to the present approach is a $4\pi$ BGO calorimeter operated by the ETH Zurich group~\cite{crivelli2006,gendotti}, which reported the current best limit on the branching ratio, $4.3\times10^{-7}$~\cite{ethresult,eth18,eth20}. These searches have gradually tightened the limit on invisible positronium decay, and a further improvement of roughly an order of magnitude would help constrain the corresponding hidden-sector models. KAPAE Phase II follows the same $4\pi$ BGO calorimetric strategy and adds several features: low-temperature operation, lead and copper shielding, an internal SiPM-based positron trigger, and installation 1000~m underground to suppress cosmic-ray- and environment-induced backgrounds. With these features, KAPAE Phase II is expected to reach a branching-ratio sensitivity of about $10^{-8}$ for the total invisible decay channel, roughly an order of magnitude beyond the current best limit.

This paper describes the design, optimization, assembly, and underground implementation of the KAPAE Phase II detector, focusing on the instrumentation: the detector concept, Geant4-based geometry optimization, positron trigger configuration, BGO scintillator performance, SiPM-based readout, the data acquisition system, low-temperature operation, and the comparison of surface and underground detector performance.

\section{Design Requirements and Detector Concept}

\subsection{Physics-driven detector requirements}

The KAPAE Phase II detector uses a compact calorimetric approach to search for total and partial invisible decays of positronium. In this measurement, a positron emitted from the radioactive source is tagged first, and the subsequent annihilation photons are measured with high detection efficiency. A candidate event for a total invisible decay is a valid positron trigger with no corresponding energy deposit in the surrounding calorimeter. A partially invisible decay instead appears as a positron-tagged event with an anomalous single-photon energy deposit, depending on the mass of the undetected particle.

These experimental signatures impose several requirements on the detector. First, the inactive volume around the positronium formation region must be minimized because ordinary annihilation photons escaping through dead regions can mimic missing-energy events. Second, the detector must have sufficient stopping power for 511 keV and 1275 keV gamma rays emitted from the $^{22}$Na source and positron annihilation. Third, the positron trigger must be located close to the source region while introducing only a small amount of passive material. Finally, the detector response must remain stable over long data-taking periods, since the expected signal rate for invisible decay is extremely small.

The KAPAE Phase I detector~\cite{KAPAE1jinst} was optimized mainly for visible multi-photon decay studies, for which fine segmentation and angular information were needed. Such a configuration, however, is not well suited to an invisible new particle search: the large number of detector elements and readout channels increases the uninstrumented space and complicates calibration. The Phase II detector was therefore redesigned around different priorities: photon containment, reduced dead area, simpler readout, and stable operation under low-background conditions.

BGO was chosen as the main scintillation material for this purpose. Its high density and high effective atomic number are suited to detecting sub-MeV to MeV gamma rays in a compact volume, as demonstrated in other BGO-based calorimeters~\cite{ishikawa}. Its scintillation decay time is relatively long compared with faster scintillators, but this is not a limitation for the present calorimetric measurement, where stopping power and stable energy measurement matter more than fast timing. Cryogenic operation reduces SiPM-related noise and also increases the BGO light yield, improving the energy resolution~\cite{BGOtemp}. The detector was therefore designed to operate underground and at cryogenic temperature. The underground location suppresses cosmic-ray-induced backgrounds, while the lead and copper shielding around the detector (Section~\ref{sec:Development}) suppresses external radiation backgrounds.

\subsection{Overall detector geometry}

\begin{figure}[t]

\centering

\includegraphics[width=0.75\textwidth]{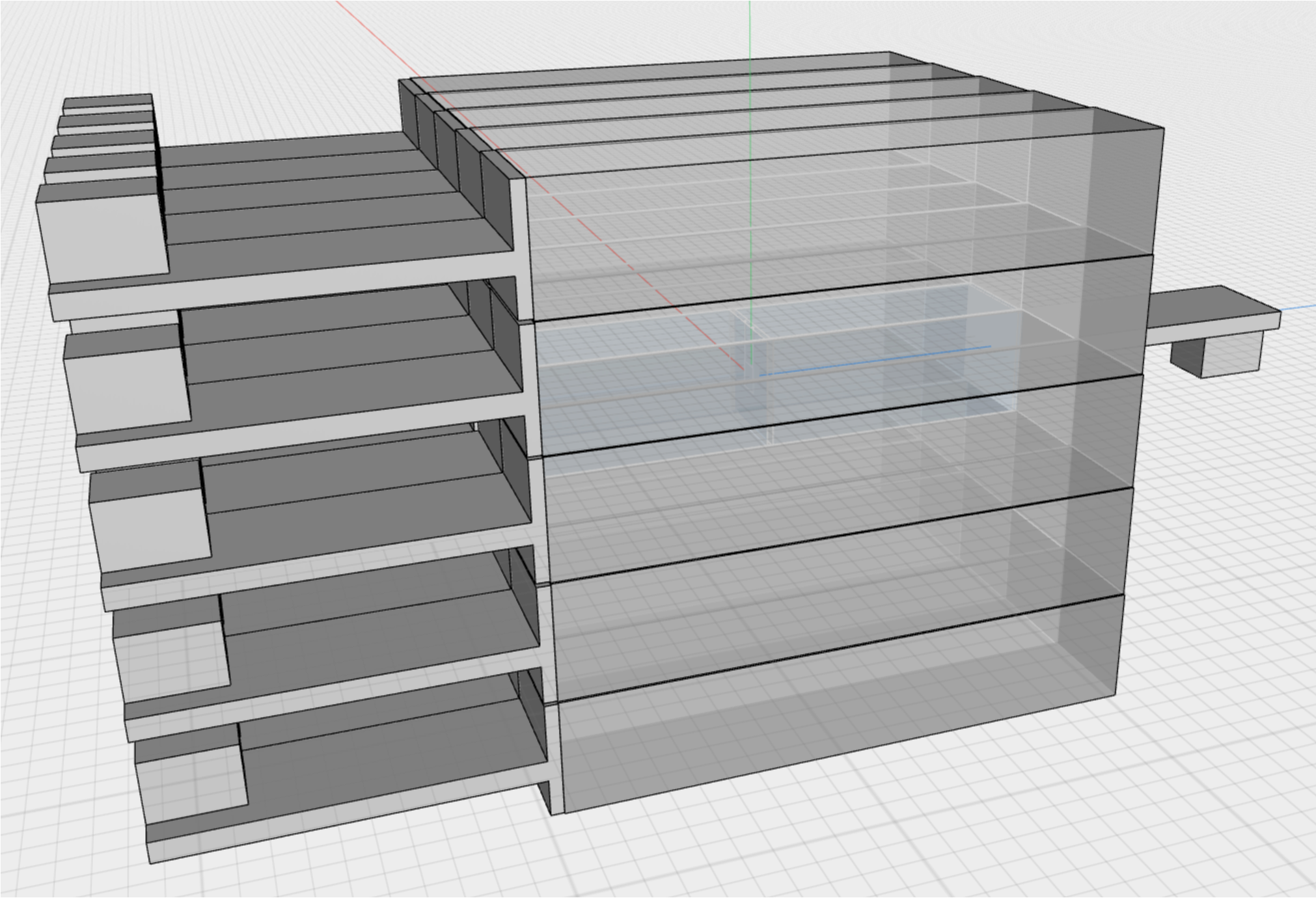}

\caption{Schematic view of the KAPAE Phase II detector geometry. The detector consists of a compact $5 \times 5$ BGO calorimeter and a central positron trigger region. The central BGO is divided into two endcap crystals to accommodate the $^{22}$Na source and PEN-film trigger.}

\label{fig:CAD}

\end{figure}

The KAPAE Phase II detector consists of a compact array of BGO scintillation crystals surrounding the central positronium formation and trigger region, as illustrated schematically in Figure~\ref{fig:CAD}. The main calorimeter is based on a $5 \times 5$ arrangement of BGO crystals. Most of the BGO crystals have dimensions of $30 \times 30 \times 150~\mathrm{mm}^{3}$, providing sufficient volume for gamma-ray absorption while keeping the detector compact.

The central position of the array is reserved for the positron source and trigger assembly. To accommodate this trigger region, the central BGO crystal is divided into two shorter endcap BGO crystals, each with dimensions of $30 \times 30 \times 75~\mathrm{mm}^{3}$. These two endcap crystals are placed on opposite sides of the source and trigger region. This configuration allows the detector to preserve gamma-ray coverage around the center while providing mechanical space for the positron trigger. The distance between the two endcap crystals can also be adjusted, which allows modifications of the source and trigger configuration without changing the entire calorimeter structure.

This geometry represents a compromise between gamma-ray containment and mechanical accessibility. A fully closed calorimeter would be preferable in terms of detection efficiency, but the detector must also include a source holder, a positron trigger, optical reflectors, SiPM readout boards, and mechanical supports. The Phase II design therefore uses a compact BGO arrangement with a small but controllable central region, so that the detector can maintain high photon coverage while allowing stable installation and maintenance of the trigger system.

The reduced segmentation, compared with the Phase I detector, also reduces the number of optical interfaces and readout channels. Each additional gap, reflector boundary, or inactive mechanical component is a possible path through which annihilation photons escape or deposit only partial energy. Fewer such paths mean fewer ordinary annihilation events are misreconstructed as missing-energy-like background; this was the guiding criterion for the Phase II geometry.

\subsection{Positron trigger concept}

The positron trigger is central to the KAPAE Phase II detector, since candidate events for the invisible new particle search are defined with respect to a tagged positron. The trigger must identify positrons emitted from the $^{22}$Na source while adding as little passive material near the source as possible. For this purpose, a thin polyethylene naphthalate (PEN) film is used as the scintillation trigger, following the same approach as the KAPAE Phase I detector~\cite{KAPAE1op}. PEN suits this role because it can act both as a thin mechanical film and as a scintillator for charged-particle detection, as previously demonstrated in a BGO-PEN phoswich detector for radiation-type discrimination~\cite{phoswich}.

The $^{22}$Na source emits a positron together with a prompt 1275 keV gamma ray. After the positron loses energy and forms positronium or annihilates directly, 511 keV annihilation photons are emitted in ordinary decay channels. In the KAPAE Phase II detector, the PEN film provides the positron tag, while the surrounding BGO crystals measure the prompt and annihilation gamma rays. This coincidence information is used to select events associated with the source decay and to reject unrelated environmental backgrounds.

The trigger is placed in the central region between the two endcap BGO crystals. This arrangement has two advantages. First, the positron can be tagged close to its emission point, which increases the trigger efficiency. Second, the surrounding BGO crystals can measure gamma rays over a large solid angle, reducing the probability that ordinary annihilation events are misidentified as candidates for the invisible new particle search. The compact placement of the trigger also reduces position-related uncertainties in the comparison between data and Monte Carlo simulations.

A thin trigger is preferred because excessive trigger material can perturb the gamma-ray field and increase secondary interactions near the source. At the same time, the trigger must be thick enough to provide a stable positron signal. Therefore, the PEN trigger configuration was treated as one of the detector parameters to be optimized experimentally and through simulation. The final trigger concept was chosen to balance trigger efficiency, low material budget, mechanical stability, and compatibility with the BGO calorimeter geometry.

The overall detector concept can be summarized as a positron-tagged compact calorimeter: the PEN film marks the start of the event, and the BGO array records whether the expected gamma-ray energy was deposited. This concept is well suited to an invisible new particle search, since the main experimental observable is simply whether a valid positron trigger is matched by the expected gamma-ray energy in the calorimeter.

\section{Geant4 Simulation for Detector Optimization}
\label{sec:Geant4}

\subsection{Detector geometry}

A Geant4~\cite{geant4} Monte Carlo simulation was developed to optimize the geometry of the KAPAE Phase II detector and to evaluate the detector response to gamma rays from the $^{22}$Na source and positron annihilation. The simulation was used as a design tool before detector assembly and later as a reference model for comparison with experimental data. The main quantities considered in the optimization were the gamma-ray containment efficiency, the effect of inactive regions between BGO crystals, and the influence of the positron trigger configuration on the measured energy spectrum.

\begin{figure}[t]

\centering

\includegraphics[width=0.85\textwidth]{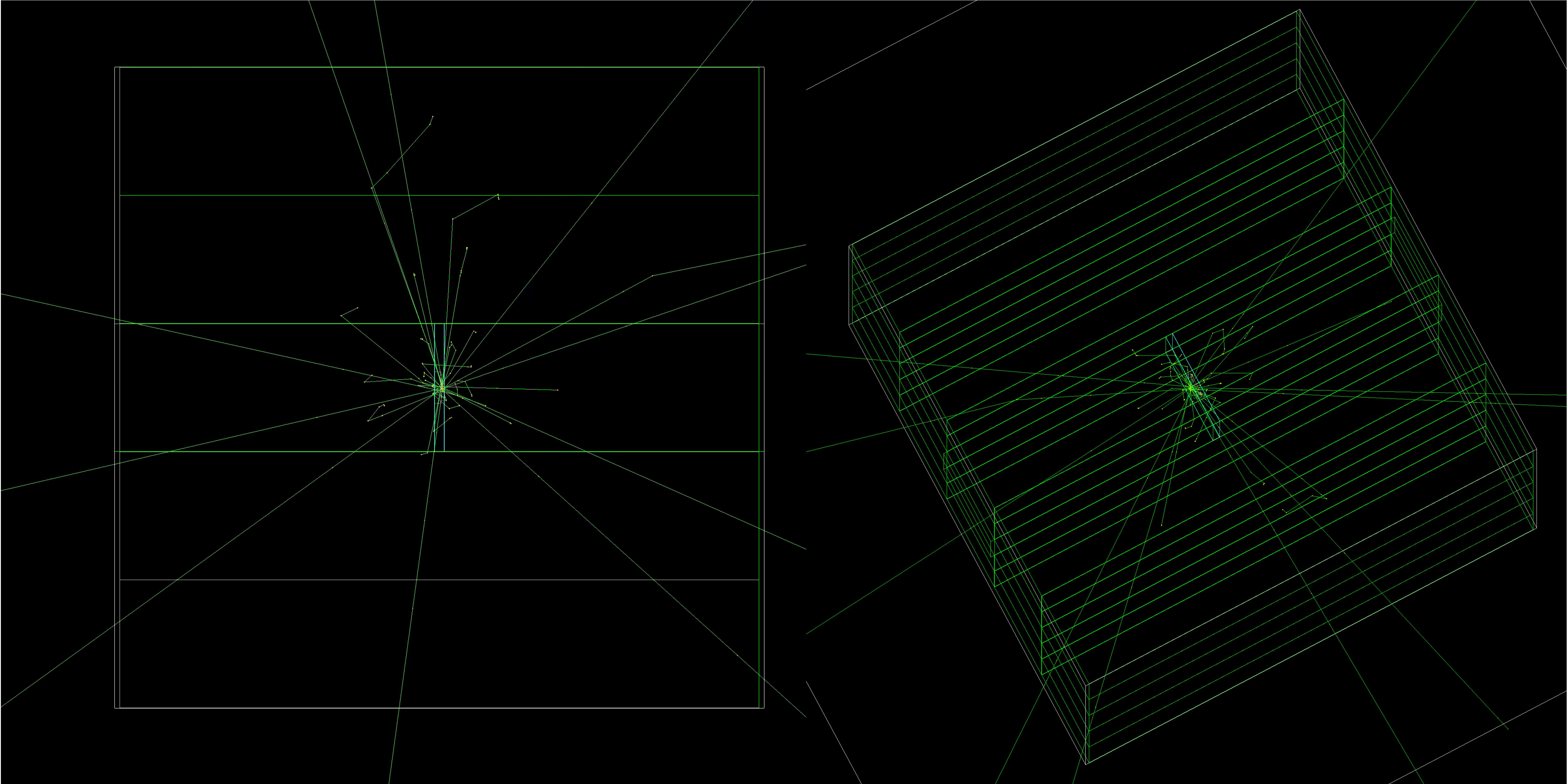}

\caption{Geant4 simulation geometry of the KAPAE Phase II detector. The model includes the compact $5 \times 5$ BGO calorimeter, two shortened endcap BGO crystals, and the central $^{22}$Na source and PEN-film trigger region.}

\label{fig:geant4_geometry}

\end{figure}

The simulated detector geometry, shown in Figure~\ref{fig:geant4_geometry}, reproduced the main components of the KAPAE Phase II detector, including the $5 \times 5$ BGO calorimeter, the two shortened endcap BGO crystals, the central source and trigger region, optical reflectors, and surrounding mechanical structures. The BGO crystals were implemented with the same dimensions as the constructed detector. The full-size BGO crystals had dimensions of $30 \times 30 \times 150~\mathrm{mm}^{3}$, while the two endcap crystals had dimensions of $30 \times 30 \times 75~\mathrm{mm}^{3}$. The $^{22}$Na source was placed at the center of the detector, between the two endcap crystals, together with the PEN-film positron trigger.

The simulation was performed using Geant4 version 10.07. The electromagnetic and radioactive decay processes were modeled using the \texttt{G4EmLivermorePhysics}, \texttt{G4RadioactiveDecayPhysics}, and \texttt{G4DecayPhysics} physics lists. These physics lists were selected because the detector response is dominated by low-energy electromagnetic interactions in the sub-MeV to MeV range, including photoelectric absorption, Compton scattering, gamma-ray escape, and radioactive decay of the $^{22}$Na source.

The $^{22}$Na decay was modeled by including the prompt 1275 keV gamma ray and positron emission. The subsequent annihilation photons were then tracked in the BGO calorimeter. For ordinary para-positronium-like two-photon annihilation, two 511 keV gamma rays were generated in approximately opposite directions from the source region. The deposited energy in each BGO channel was recorded event by event. This allowed the simulation to evaluate how often ordinary annihilation events could appear as low-energy or missing-energy-like events because of photon escape, dead regions, or incomplete energy deposition.

\subsection{PEN trigger optimization}

The positron trigger configuration was also studied using Geant4 simulations. The trigger consisted of a thin poly(ethylene naphthalate) (PEN) film plastic scintillator with a thickness of \(125~\mu\mathrm{m}\) per layer. Since candidate events for the invisible new particle search are defined by the presence of a valid positron tag and the absence or reduction of gamma-ray energy deposition in the BGO calorimeter, the trigger must provide a sufficiently high positron detection efficiency while introducing minimal disturbance to gamma-ray transport. A thicker PEN trigger improves the positron signal and trigger stability, but it can also increase the probability of gamma-ray interaction or energy loss near the source. Therefore, the PEN thickness was optimized by considering both the positron trigger efficiency and gamma-ray attenuation.

\begin{figure}[t]

\centering

\includegraphics[width=0.85\textwidth]{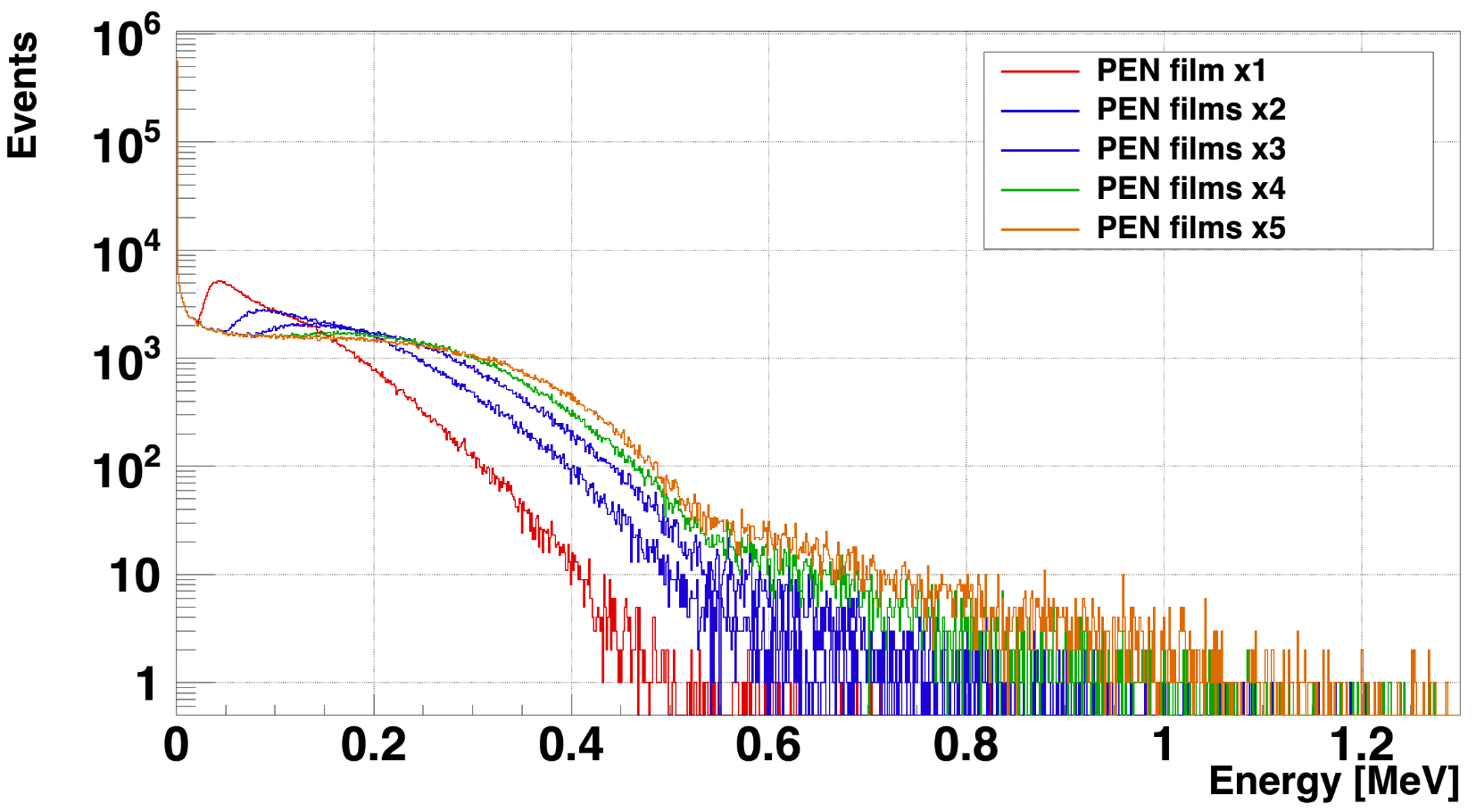}

\caption{Geant4 simulation results for the PEN-film trigger optimization. The trigger rate increases with PEN thickness and tends to saturate above approximately two layers, while gamma-ray energy-loss probabilities increase with additional trigger material. A three-layer-equivalent PEN configuration was selected as the final design.}

\label{fig:pen_trigger_optimization}

\end{figure}

Several PEN-film configurations were simulated by varying the number of \(125~\mu\mathrm{m}\)-thick layers, as summarized in Figure~\ref{fig:pen_trigger_optimization}. For each configuration, the trigger response to positrons emitted from the \(^{22}\mathrm{Na}\) source was evaluated together with the energy-loss probability for gamma rays passing through the trigger region. The simulated trigger rate increased with the total PEN thickness and tended to saturate above approximately two layers, corresponding to a total thickness of \(250~\mu\mathrm{m}\). This behavior indicates that additional PEN material beyond this thickness provides only a limited increase in the positron trigger efficiency.

The effect of the PEN trigger on gamma rays was evaluated separately for low-energy photons and for \(511~\mathrm{keV}\) annihilation photons, since the partial invisible new particle search is sensitive to a broad photon-energy region, while the ordinary annihilation background is dominated by \(511~\mathrm{keV}\) photons. The simulation showed that increasing the number of PEN layers improves the positron signal but also increases the probability of gamma-ray energy loss in the trigger material. The final configuration was therefore selected as a compromise between adequate trigger efficiency and minimal gamma-ray attenuation.

Based on these studies, three layers of \(125~\mu\mathrm{m}\)-thick PEN film, corresponding to a total thickness of \(375~\mu\mathrm{m}\), were adopted for the final detector configuration. This configuration provides a stable positron trigger signal and allows a sufficiently high trigger threshold to reject electronic noise and gamma-induced background in the trigger channel. At the same time, the amount of material remains small compared with that of the surrounding BGO calorimeter, thereby limiting the influence of the trigger on the gamma-ray energy measurement.

The optimized PEN trigger was implemented in the central detector region between the two endcap BGO crystals. In the final event-selection scheme, the PEN signal defines the positron-tagged event, while the BGO calorimeter measures the prompt \(1275~\mathrm{keV}\) gamma ray and the annihilation photons. This trigger--calorimeter combination provides the basic experimental condition for identifying ordinary annihilation events and searching for missing-energy-like signatures.

The geometry and trigger optimization, together with the BGO crystal and readout improvements described in the following sections, give the KAPAE Phase II detector an improved energy resolution relative to the crystals used in earlier BGO-based searches (Section~\ref{sec:Performance}). Based on Geant4 sensitivity studies with this configuration, the detector is expected to reach a branching-ratio sensitivity of about $10^{-8}$ for the total invisible decay channel, roughly an order of magnitude beyond the current best reported limit of $4.3 \times 10^{-7}$~\cite{ethresult}.

\section{BGO Crystal Performance}
\label{sec:BGO}

\begin{figure}[t]

\centering

\includegraphics[width=0.5\textwidth]{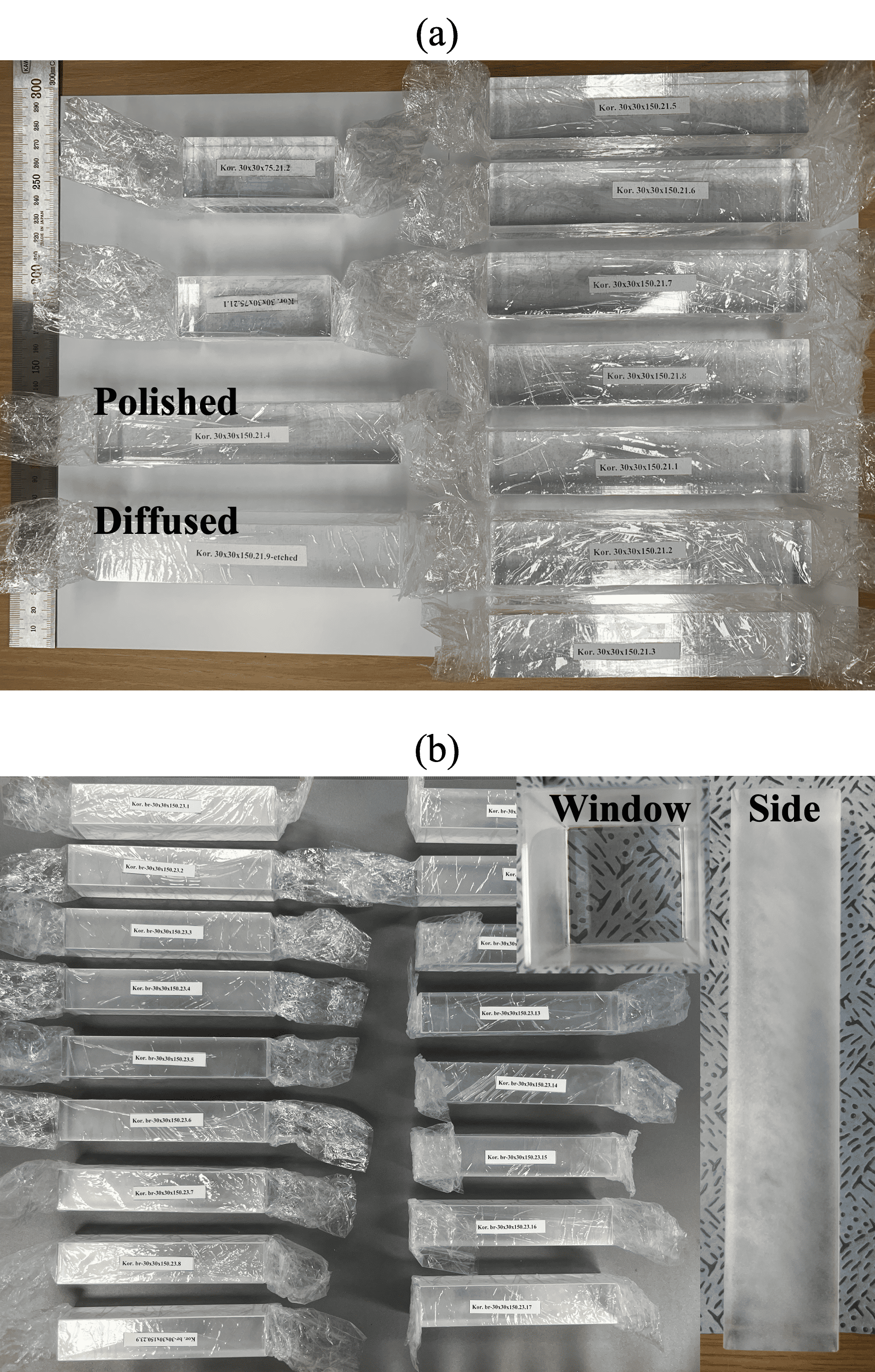}

\caption{Photographs of the BGO scintillation crystals used in the KAPAE Phase II detector. (a) Crystals with dimensions of $30 \times 30 \times 150~\mathrm{mm}^{3}$ and $30 \times 30 \times 75~\mathrm{mm}^{3}$, with polished and diffused surface finishes. (b) Detail of a diffused crystal, with a close-up view of the exit window and side surface.}

\label{fig:BGO_crystals}

\end{figure}

The BGO crystals used for the KAPAE Phase II calorimeter were custom-fabricated by the Nikolaev Institute of Inorganic Chemistry, Siberian Branch of the Russian Academy of Sciences (NIIC SB RAS). Two surface treatments, polished and diffused, were compared for their optical performance. Figure~\ref{fig:BGO_crystals} shows photographs of the delivered crystals, including the exit window and side-surface treatment of a diffused sample. In our tests, the diffused surface gave an average of about 22\% higher light yield than the polished surface, with individual crystals ranging from 16.6\% to 29.4\%~\cite{diffuse}. This is consistent with Geant4 optical-transport simulations, which attribute the improvement to a higher probability of photon extraction at the roughened surface, where total internal reflection is reduced. The diffused treatment was therefore adopted for the crystals used in the final detector.

\subsection{Temperature dependence of scintillation response}

Since the KAPAE Phase II detector was designed to operate in an underground cryogenic environment, the temperature dependence of the BGO scintillation response had to be evaluated beforehand. The scintillation decay time of BGO changes with temperature, and this affects the waveform shape, the charge integration window, pile-up behavior, and the FADC readout configuration.

The temperature dependence was measured using a BGO crystal optically coupled to a custom SiPM array. The SiPM signals were read out by the NKFADC500 system, together with a dedicated SiPM bias supply board. During the measurement, the detector temperature was monitored near the BGO assembly. Since the BGO crystal was enclosed in a mechanical structure, the measured temperature did not exactly correspond to the internal crystal temperature. The uncertainty associated with this temperature difference was estimated to be approximately $\pm 2.5^{\circ}\mathrm{C}$.

\begin{figure}[t]

\centering

\includegraphics[width=0.80\textwidth]{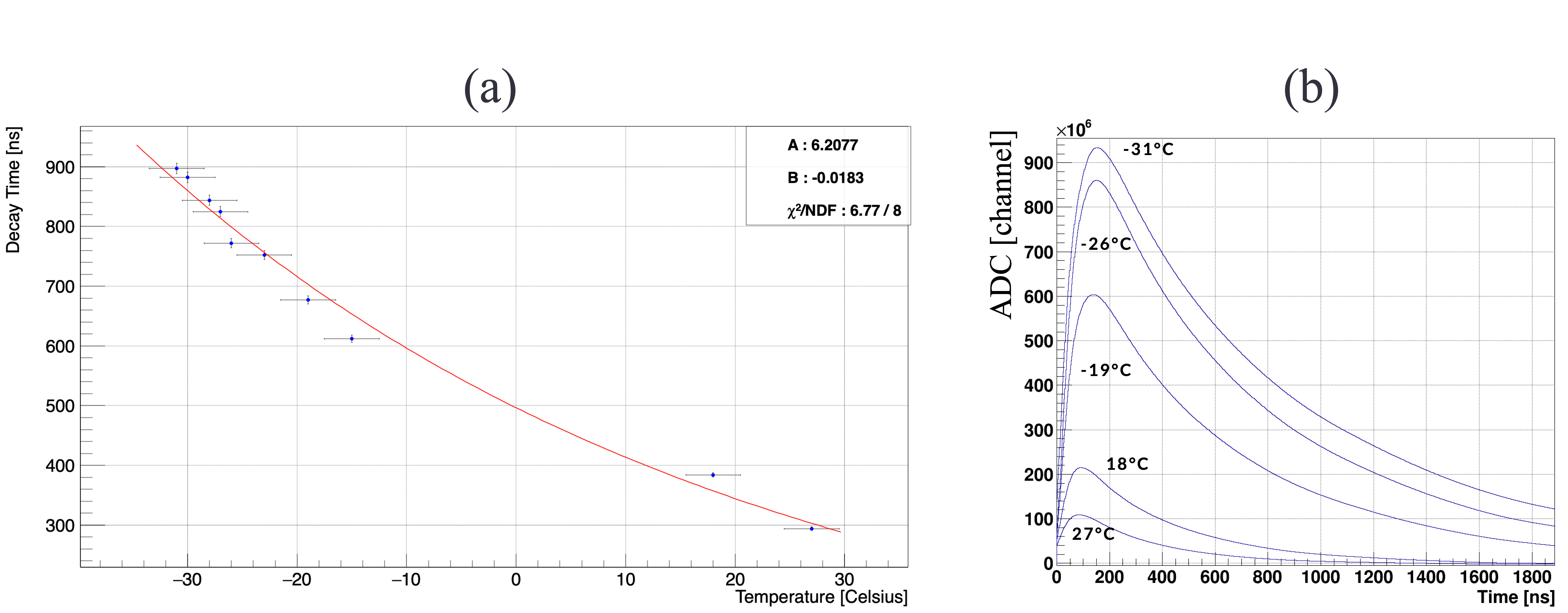}

\caption{Temperature dependence of the BGO scintillation response. (a) Decay time of the BGO scintillation crystal as a function of temperature. (b) Average BGO scintillation waveforms measured at different temperatures.}

\label{fig:BGO_temp_decay}

\end{figure}

The measured BGO waveform became slower as the temperature decreased, as shown in Figure~\ref{fig:BGO_temp_decay}. The scintillation decay time increased from approximately 300 ns at room temperature to 789 ns at about $-26^{\circ}\mathrm{C}$. The temperature dependence was fitted with an exponential function,

\begin{equation}
\tau(T) = 493 \exp\left(-0.018T\right)~\mathrm{ns}.
\label{eq:bgo_decay_temperature}
\end{equation}

where $\tau$ is the BGO scintillation decay time in ns and $T$ is the temperature in $^{\circ}\mathrm{C}$. This behavior is consistent with previously reported temperature-dependent properties of BGO~\cite{BGOtemp}. From this relation, the decay time at the actual detector operating temperature of approximately $-37^{\circ}\mathrm{C}$ is expected to be about 960 ns.

This increase in decay time affects the data acquisition system directly. If the charge integration window is too short, part of the scintillation light is lost at low temperature, which shifts the reconstructed energy and degrades the energy resolution. The FADC waveform acquisition and charge integration conditions were therefore adjusted to match the longer BGO pulse shape under cryogenic operation. Cryogenic operation reduces SiPM-related noise, but it requires a readout configuration compatible with the slower BGO scintillation signal.

The temperature study set the waveform length, integration window, and pedestal treatment used in the later data-taking period. This allowed the detector response measured underground to be compared consistently with room-temperature calibration data and with the Geant4 detector-response simulation.

\section{Photodetector and Data Acquisition System}
\label{sec:Electronics}

\subsection{SiPM array and preamplifier}

The photodetector system of the KAPAE Phase II detector was designed to provide stable light readout from large BGO crystals under compact and cryogenic operating conditions. Since the detector is installed inside a shielded underground setup, the readout components must be mechanically compact, electrically stable, and suitable for long-term operation with limited access after installation. In addition, the photosensitive area should match the $30 \times 30~\mathrm{mm}^{2}$ exit window of each BGO crystal in order to minimize optical dead area.

In the initial photodetector selection stage, a square photomultiplier tube and a custom SiPM array were compared. The Hamamatsu R11265U-300 PMT and a custom-made SiPM array based on the Hamamatsu S13360-6075 series were tested using a BGO crystal and a $^{22}$Na source. The measured energy resolution for the 511 keV peak was 16.5\% for the PMT and 15.1\% for the SiPM array. Although both devices provided acceptable readout performance, the SiPM-based solution was selected because it offers a compact geometry, low-voltage operation, and easier integration with a multi-channel calorimeter.

The active-area coverage was another important factor in the selection. The active window of the PMT was smaller than the $30 \times 30~\mathrm{mm}^{2}$ BGO crystal surface, which introduced an optical mismatch. The initial $3 \times 3$ SiPM array also had a limited effective active area and relatively large inactive gaps between individual SiPM elements. To overcome this limitation, an improved $4 \times 4$ SiPM array was developed. This configuration increased the photosensitive coverage and allowed the array size to match the BGO crystal window more closely.

\begin{figure}[t]

\centering

\includegraphics[width=0.85\textwidth]{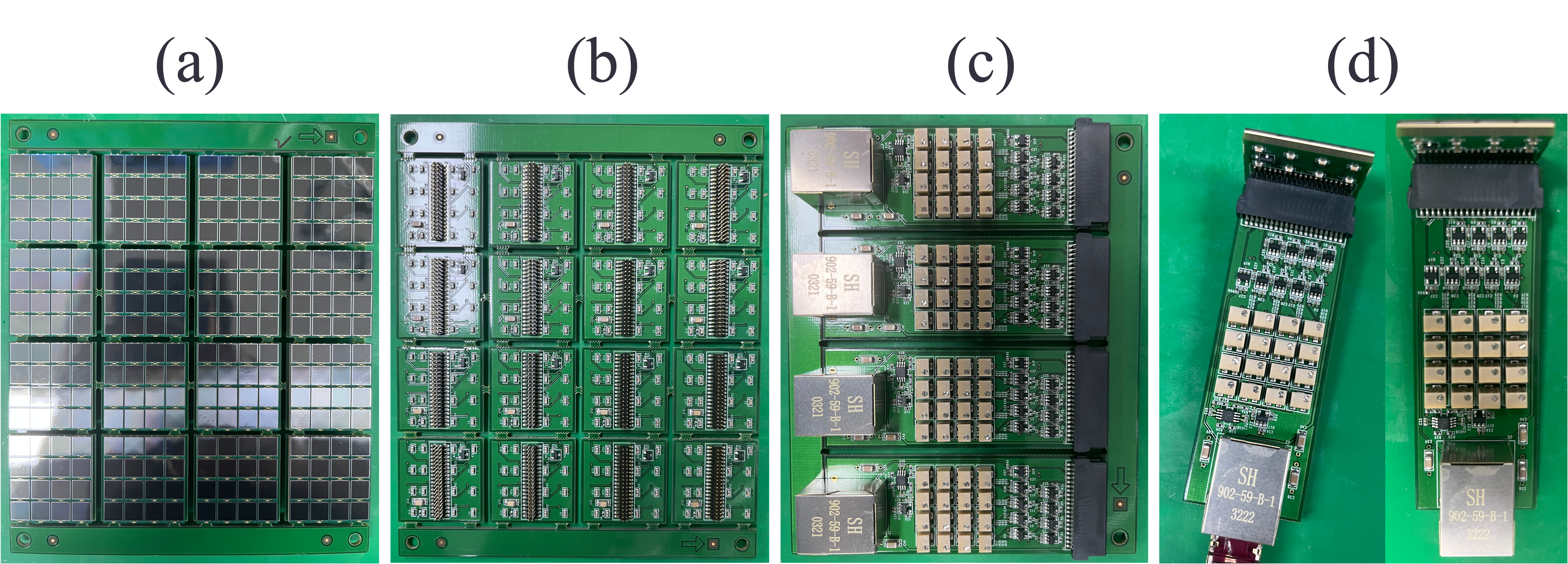}

\caption{Photographs of the final SiPM array and preamplifier module. (a) Front view and (b) rear view of the final SiPM array board. (c) Preamplifier board. (d) Integrated SiPM array and preamplifier modules.}

\label{fig:SiPM}

\end{figure}

Each detection channel consists of sixteen SiPM elements arranged in a $4 \times 4$ array and coupled to a dedicated preamplifier board, developed in collaboration with Notice Co., Ltd.~\cite{Notice}, as shown in Figure~\ref{fig:SiPM}. Since the gain of individual SiPM elements can differ, adjustable resistors were implemented for gain balancing. This design reduces the degradation of energy resolution caused by nonuniform SiPM gains within one readout channel. The SiPM array and preamplifier board were designed as an integrated module to reduce noise pickup and simplify detector assembly.

During early tests, the electronic noise increased with the cable length between the SiPM array and the preamplifier. This is a concern for the underground detector, where long cabling and compact shielding structures are unavoidable. To reduce this noise, the preamplifier board was directly coupled to the SiPM array, which shortened the sensitive analog signal path before amplification and improved the stability of the readout chain.

Category-6 LAN cables and connectors were adopted for signal transmission and system integration, simplifying the cabling between detector modules and the DAQ system while giving a practical multi-channel connection scheme. The final SiPM and preamplifier module was designed with energy resolution, mechanical stability, low-noise operation, and compatibility with the underground installation all taken into account.

\subsection{DAQ system}

A custom data acquisition system was developed and manufactured by Notice Co., Ltd. for the KAPAE Phase II detector to process multi-channel SiPM signals and provide flexible trigger control~\cite{Notice}. The system was designed to handle the BGO calorimeter channels and the PEN trigger channel under both ground-level test conditions and underground cryogenic operation. The DAQ system supplies power to the preamplifier boards, digitizes the detector signals, records event-level parameters, and provides timing information for event reconstruction.

The final DAQ system can process analog signals from up to 56 photodetector channels. Each channel is digitized using a 12-bit analog-to-digital converter with a sampling frequency of 62.5 MHz. This sampling rate is sufficient for measuring the relatively slow BGO scintillation waveform, especially under low-temperature conditions where the BGO decay time becomes longer. The system also includes a dedicated time-to-digital converter, providing a trigger timing resolution of approximately 1.3 ns.

The DAQ system provides flexible coincidence trigger logic. Triggers can be formed using arbitrary combinations of selected channels, allowing the system to operate in several modes depending on the measurement purpose. For example, calibration measurements can be performed using BGO self-trigger or coincidence conditions, while the positronium measurement uses the PEN trigger as the primary event tag. This flexibility was used during detector commissioning, when the trigger logic had to be adjusted after comparing ground-level and underground data.

For each triggered event, the DAQ records not only the integrated charge but also several waveform-related parameters. These include the TDC trigger time, total charge, peak time, pulse height, mean time, and temperature information for each channel. The pulse height is defined from the ADC value at the signal maximum, while the mean time is calculated from the charge-weighted timing information after the pulse peak. These parameters are useful for identifying abnormal pulses, pile-up events, low-energy noise, and background-like events.

The system also allows waveform storage for a configurable fraction of triggered events, which is used to monitor detector stability and to validate the offline reconstruction algorithm without storing every waveform and greatly increasing the data size. For long-term underground operation, storing compact event-level quantities while keeping selected waveforms balances analysis flexibility against data volume.

The trigger logic was improved during the detector development. In the original configuration, the trigger time was determined when the integrated charge exceeded a predefined threshold. This method introduced an amplitude-dependent time walk, reaching up to approximately $1~\mu\mathrm{s}$. To reduce this effect, the trigger definition was modified to use a pulse-height threshold first, followed by a charge-sum condition. This change aligned the trigger reference more closely with the rising edge of the PEN signal and reduced the time walk to approximately 100 ns.

\subsection{Pedestal correction and charge integration}

Pedestal correction and charge integration need to be accurate for the KAPAE Phase II detector, since candidate events for the invisible new particle search are selected from events with small or missing BGO energy deposition. In this regime, even a small pedestal fluctuation can distort the reconstructed low-energy spectrum and produce artificial background. The DAQ and reconstruction procedure were therefore designed to evaluate the pedestal event by event and apply a consistent charge integration method to all channels.

The pedestal is calculated from a waveform region preceding the trigger signal. A $2~\mu\mathrm{s}$ pre-trigger window is used to estimate the baseline level and its variance. The variance is used as a quality indicator for the pedestal estimate. If the variance is less than 1 ADC count, the measured pedestal value is used directly for the charge calculation. If the variance exceeds this threshold, the pedestal value from the previous stable $2~\mu\mathrm{s}$ window is used instead. This procedure prevents local noise fluctuations or pre-trigger pulse contamination from biasing the integrated charge.

The charge of each pulse is then obtained by integrating the pedestal-subtracted ADC values within a predefined time window. A common integration window of $8~\mu\mathrm{s}$ was adopted for all channels in order to reduce channel-dependent reconstruction bias. This window length matters especially for low-temperature operation, where the BGO scintillation decay time increases to approximately 960~ns (Section~\ref{sec:BGO}). A shorter window would lose part of the scintillation signal, whereas the $8~\mu\mathrm{s}$ window includes the slow BGO component while still rejecting baseline noise and pile-up.

In addition to total charge, pulse-shape parameters such as peak time, pulse height, and mean time are stored for each event. These quantities provide information on the timing and shape of the signal and can be used to remove abnormal events. For example, events with inconsistent peak timing or unusually broad pulse shapes can be associated with pile-up, electronic noise, or nonstandard background interactions. The use of these parameters improves the reliability of event selection, especially in the low-energy region.

The pedestal correction algorithm was checked using underground data. Under low-temperature and underground conditions, the pedestal distribution remained stable, and the ADC variance was well controlled, indicating that the baseline was managed appropriately during charge integration and that the DAQ system provides a reasonably stable low-energy reconstruction. Such stability is needed for the invisible new particle search, since the main background region is sensitive to small fluctuations in the reconstructed energy.

Overall, the photodetector and DAQ system were optimized as an integrated readout chain. The custom $4 \times 4$ SiPM array improves optical coverage of the BGO crystal, the directly coupled preamplifier reduces analog noise, the custom DAQ provides flexible trigger and timing information, and the pedestal correction procedure stabilizes the reconstructed charge. These developments provide the basis for long-term underground operation of the KAPAE Phase II detector.

\begin{figure}[t]
    \centering
    \includegraphics[width=0.80\textwidth]{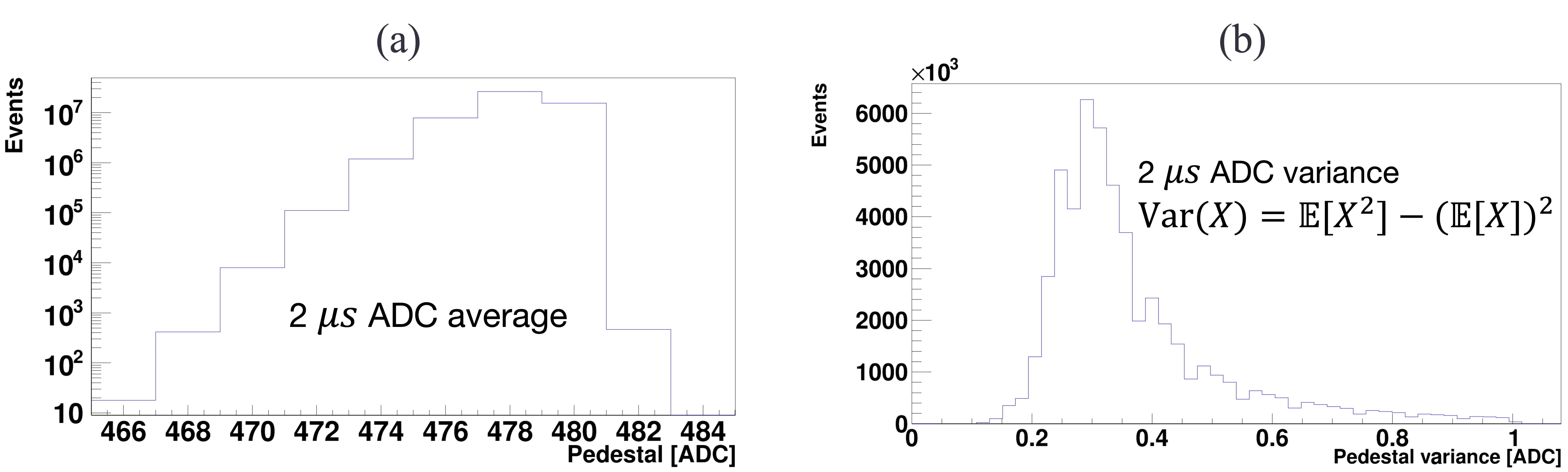}
    \caption{Pedestal distribution and ADC variance for a representative BGO channel under underground low-temperature operation. The pedestal is evaluated from a $2~\mu\mathrm{s}$ pre-trigger window, and the previous stable pedestal value is used when the variance exceeds the predefined threshold.}
    \label{fig:pedestal}
\end{figure}

\section{Detector Assembly}
\label{sec:Development}

\subsection{Detector assembly}

The KAPAE Phase II detector was assembled using the optimized configuration described in the previous sections. The initial detector concept considered SiPM readout from both ends of each BGO crystal. However, based on the measured energy resolution and the operational requirements for long-term underground measurements, the final detector adopted a single-side SiPM readout configuration. This choice reduced the number of photodetectors, preamplifier channels, cables, and calibration constants, while maintaining sufficient energy resolution for the calorimetric measurement.

The detector consists of a compact BGO array, SiPM-preamplifier modules, optical reflectors, mechanical support structures, and the central positron source and PEN-film trigger assembly. Each BGO crystal was wrapped with enhanced specular reflector film to improve optical photon collection. The SiPM array was optically coupled to one end of each BGO crystal, and the coupled module was fixed in the detector frame to maintain stable optical contact during cooling and long-term operation.

The central region of the detector was assembled with special attention to the source and trigger geometry. The $^{22}$Na source and PEN-film positron trigger were placed between the two shortened endcap BGO crystals. This configuration allowed positrons emitted from the source to be tagged close to the emission point while maintaining gamma-ray coverage around the central region. The mechanical structure was designed so that the trigger position and the spacing between the two endcap crystals could be controlled reproducibly.

After the assembly of the BGO and SiPM modules, the channel mapping was established for the full detector. The channel map defines the correspondence between each BGO crystal, SiPM-preamplifier module, DAQ input channel, and geometrical position in the detector. This mapping is used for channel-by-channel calibration, event reconstruction, and comparison with the Geant4 simulation. The same channel definition was used consistently for ground-level measurements, underground operation, and simulation-data comparison.

\begin{figure}[t]

\centering

\includegraphics[width=0.75\textwidth]{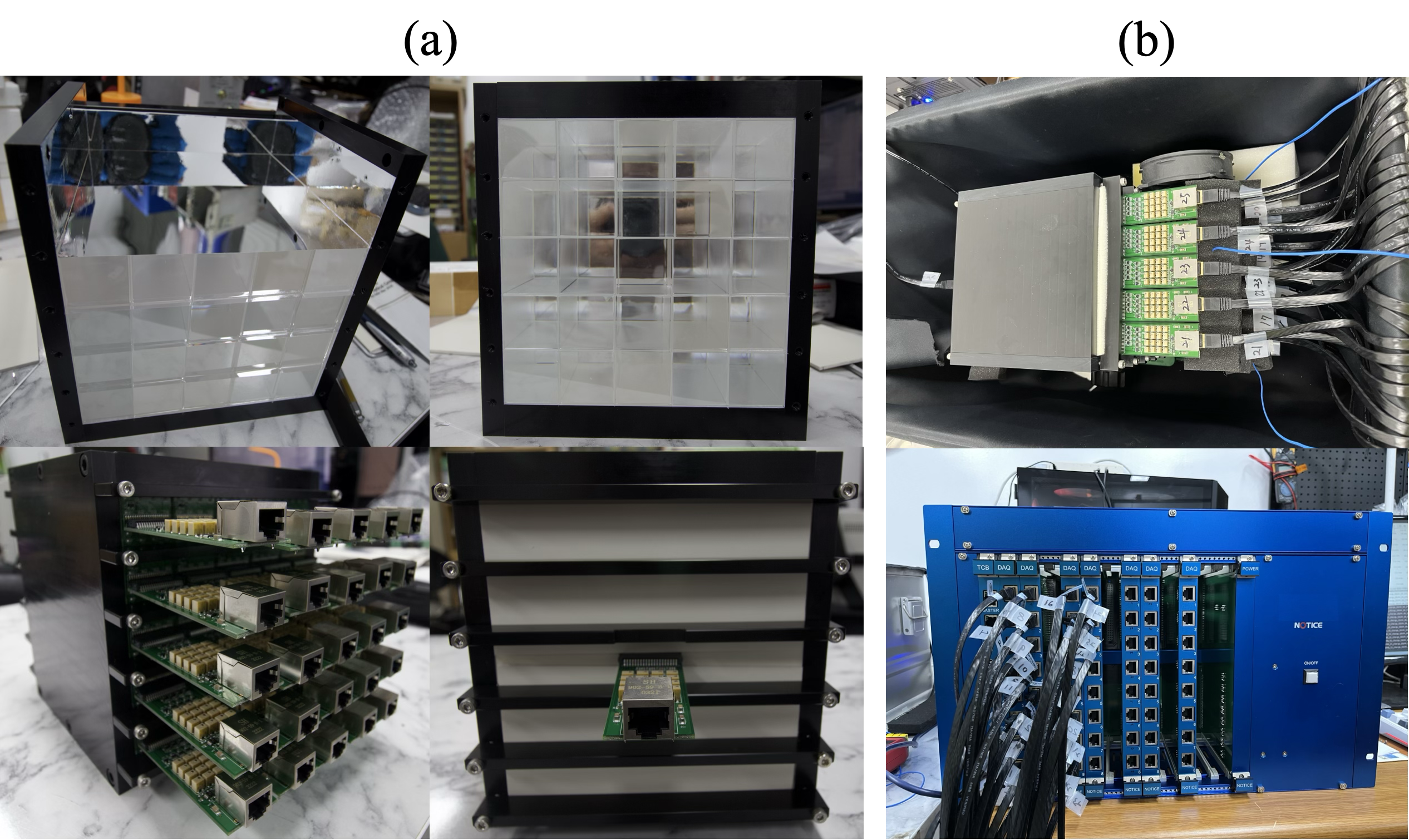}

\caption{Assembled KAPAE Phase II detector. (a) Rear view of the BGO array with the SiPM-preamplifier modules mounted on the crystal surface prior to housing. (b) Front view of the array, with the central positron source and trigger region visible. (c) SiPM-preamplifier modules and readout cabling attached to the assembled array.}

\label{fig:detector_assembly}

\end{figure}

\begin{figure}[t]

\centering

\includegraphics[width=0.75\textwidth]{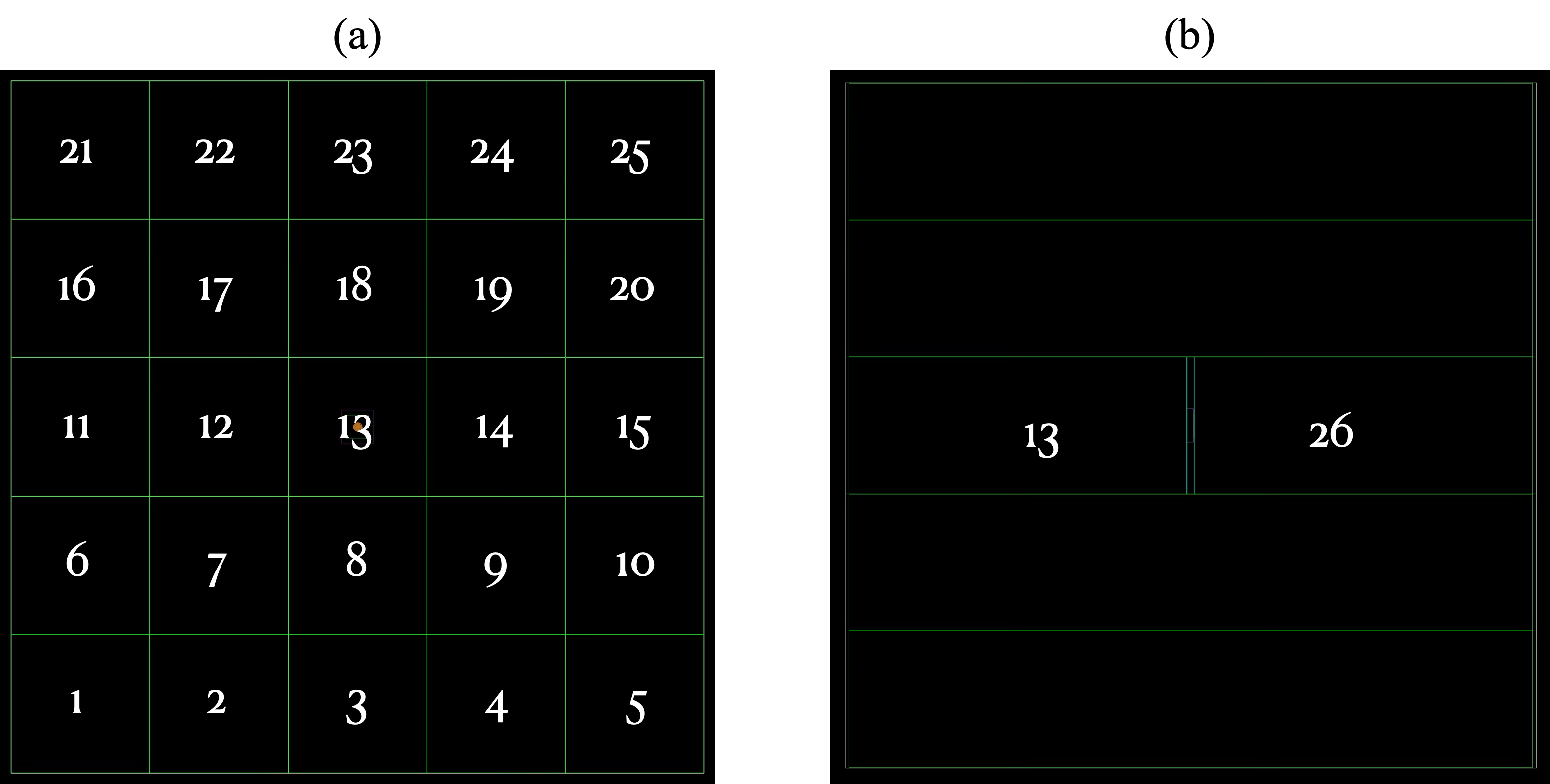}

\caption{Channel numbering scheme of the KAPAE Phase II detector. (a) Numbering of the $5 \times 5$ BGO array, viewed from the front. (b) Numbering of the two endcap BGO crystals (channels 13 and 26) at the central source and trigger position.}

\label{fig:channel_mapping}

\end{figure}

Before underground installation, the assembled detector was tested at ground level and room temperature. These tests were used to verify the operation of all readout channels, check the trigger response, confirm the DAQ configuration, and obtain initial gamma-ray spectra from the $^{22}$Na source. The ground-level measurements also provided a reference dataset for evaluating the improvement obtained after underground and cryogenic operation. Figure~\ref{fig:detector_assembly} shows the assembled $5 \times 5$ BGO array with the SiPM-preamplifier modules attached, and Figure~\ref{fig:channel_mapping} shows the channel numbering scheme adopted for the array and for the two endcap crystals.

\subsection{Low-temperature chamber and shielding}

The detector was designed to operate inside a cryogenic shielding system in order to suppress both environmental background and SiPM-related noise. The shielding system consists of an inner oxygen-free copper chamber and outer lead shielding. The copper chamber provides a clean inner volume around the detector and reduces background from lead fluorescence and nearby materials, while the lead shielding suppresses external gamma-ray backgrounds.

The shielding thickness was designed to be at least $100~\mathrm{mm}$ of lead and $10~\mathrm{mm}$ of copper in all directions. This thickness was selected to reduce environmental gamma-ray backgrounds while remaining compatible with the available space inside the cryogenic chamber. The total size of the copper chamber and lead shielding was constrained by the inner dimensions of the cooling system. During the final assembly, a small lateral gap was required because of mechanical interference between the lead blocks and the cryogenic chamber structure. To compensate for this gap, an additional external shielding layer was installed on the corresponding side. Figure~\ref{fig:shielding_installation}(a) shows the exploded view of the shielding design, together with the oxygen-free copper chamber and the lead shielding blocks.

The mechanical load of the lead shielding was also considered in the chamber design. Since the lead blocks placed above the copper case exert a large load on the detector housing, a $6~\mathrm{mm}$-thick stainless-steel plate was installed on the top surface of the copper chamber. It prevents mechanical deformation of the inner chamber and helps maintain the alignment of the detector during cooling and long-term operation.

\begin{figure}[t]

\centering

\includegraphics[width=0.75\textwidth]{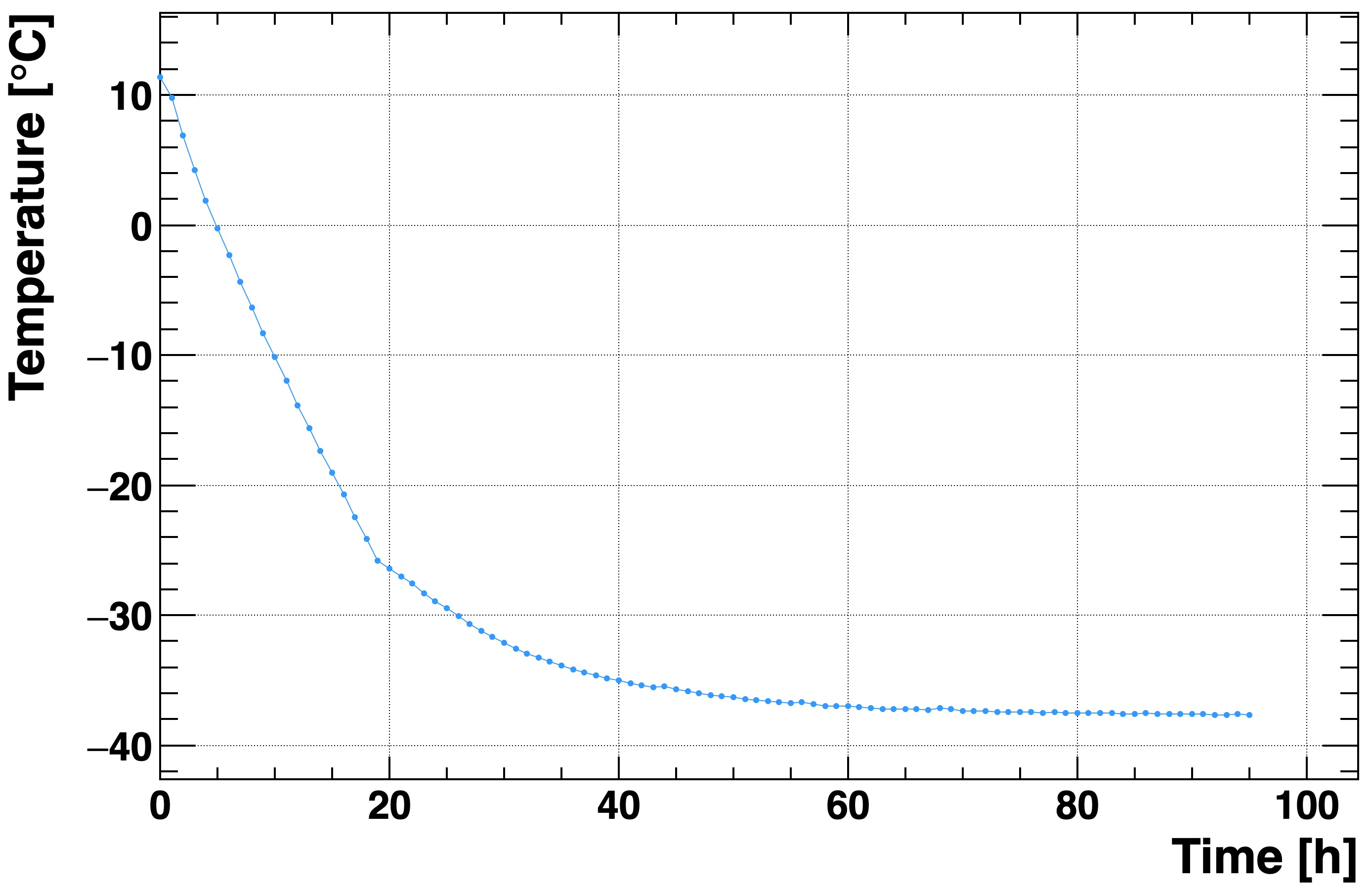}

\caption{Temperature evolution measured near the SiPM preamplifier inside the copper chamber after the cryogenic system was started. The system approached the operating temperature after approximately three days and remained stable at about $-37.5^{\circ}\mathrm{C}$ with a fluctuation of approximately $\pm 0.5^{\circ}\mathrm{C}$.}

\label{fig:temperature_stability}

\end{figure}

The cryogenic system was used to lower the detector temperature and stabilize the SiPM-based readout. The temperature was monitored near the SiPM preamplifier inside the copper chamber. After the cooling system was started, the internal temperature reached the operating setpoint after approximately three days, as shown in Figure~\ref{fig:temperature_stability}. The detector was then stabilized at about $-37.5^{\circ}\mathrm{C}$ with a fluctuation of approximately $\pm 0.5^{\circ}\mathrm{C}$ for 19 consecutive days.

Thermal stability matters for the KAPAE Phase II detector because the BGO light output, decay time, and SiPM gain are all temperature dependent. A stable temperature reduces gain drift and improves the reproducibility of the charge reconstruction. Cryogenic operation also reduces SiPM dark noise, which matters for the low-energy region relevant to the invisible new particle search. The low-temperature chamber and shielding system thus provide both background suppression and readout stabilization.

\subsection{1000~m deep Yemi underground installation}

The final detector system was installed in the Yemi underground laboratory for low-background operation. The laboratory suppresses the cosmic-ray muon flux by a factor of over 330,000 compared with the surface~\cite{Yemiconst,Yemicosine}, which suppresses high-energy background events in the BGO calorimeter. The complete installation included the assembled detector, copper chamber, lead shielding, cryogenic chamber, DAQ system, power supplies, slow-control components, and data storage system.

The installation procedure was carried out in several steps. First, the assembled detector was placed inside the copper chamber with the source and trigger assembly fixed at the central position. The SiPM-preamplifier modules were connected to the DAQ system through the prepared cabling scheme. The copper chamber was then closed and surrounded by lead shielding. After the shielding structure was completed, the cryogenic chamber was operated to bring the detector to the target low-temperature condition.

\begin{figure}[t]

\centering

\includegraphics[width=0.85\textwidth]{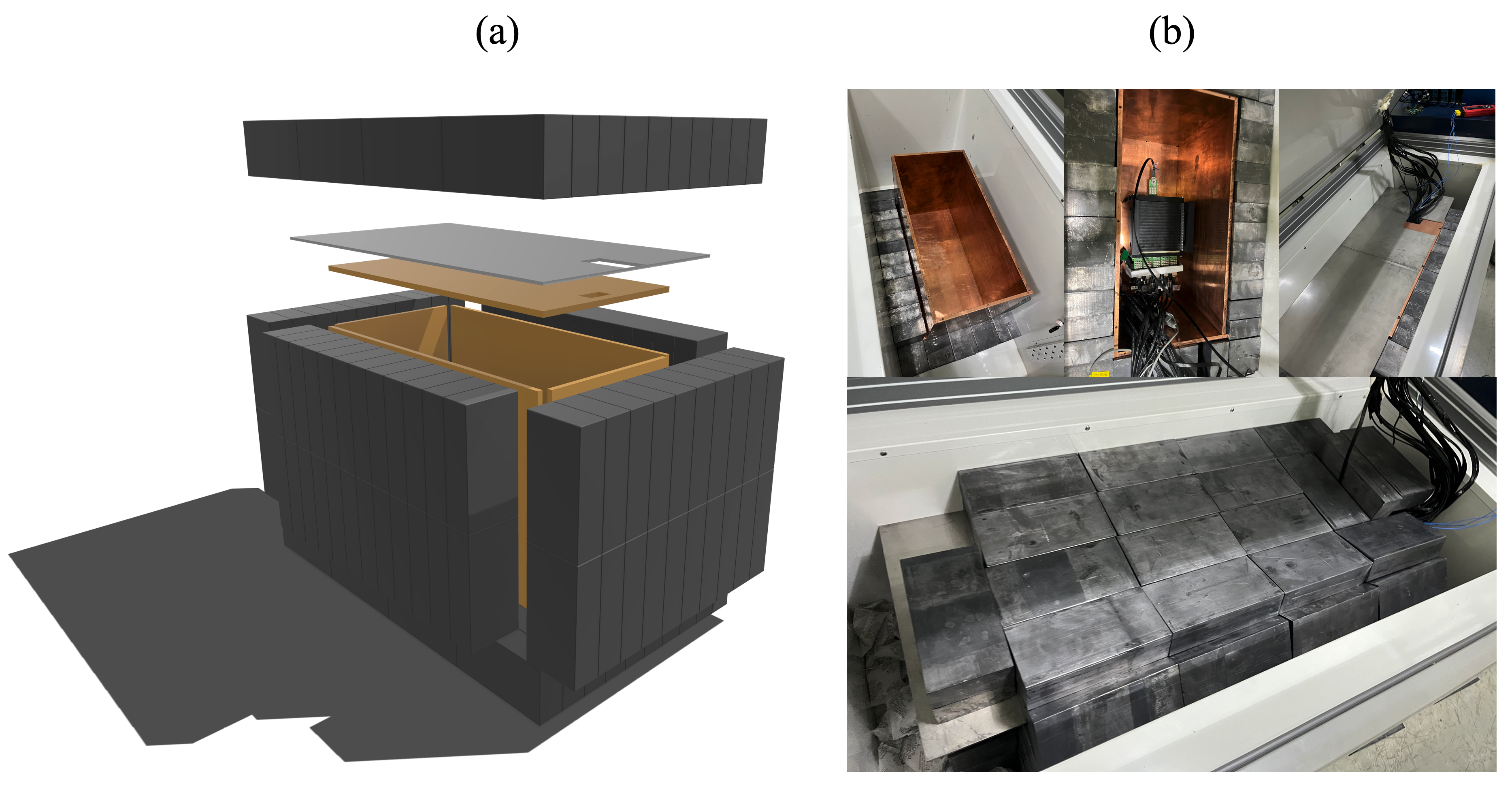}

\caption{Shielding structure and underground installation of the KAPAE Phase II detector. (a) Exploded CAD view of the cryogenic shielding, showing the lead shielding blocks (gray), the oxygen-free copper chamber (gold), and the $6~\mathrm{mm}$-thick stainless-steel plate. (b) Photographs taken during installation at the Yemi underground laboratory: the empty copper chamber before closure and the chamber with the SiPM-preamplifier modules and cabling installed (top), and the completed lead shielding stacked around the copper chamber with the readout cables routed to the outside (bottom).}

\label{fig:shielding_installation}

\end{figure}

The DAQ system was located outside the shielding and connected to the detector through feedthroughs and signal cables. This layout allowed the detector volume to remain compact while keeping the readout electronics accessible for monitoring and maintenance. During underground operation, the detector temperature, trigger rate, pedestal stability, and channel response were monitored continuously. These monitoring quantities were used to confirm that the detector remained stable over long data-taking periods. Figure~\ref{fig:shielding_installation}(b) shows the assembled copper chamber and lead shielding at the installation stage, with the internal SiPM-preamplifier cabling routed through the chamber wall.

\section{Detector Performance}
\label{sec:Performance}

\subsection{Ground-level measurement}

The assembled KAPAE Phase II detector was first evaluated at ground level and room temperature before underground installation. The purpose of this measurement was to verify the operation of all BGO channels, establish the channel-by-channel calibration procedure, evaluate the trigger response, and obtain a reference dataset under unshielded environmental conditions. The ground-level measurement also provided a direct comparison point for assessing the effect of underground and cryogenic operation.

The detector was operated with the $^{22}$Na source placed at the central trigger region. The PEN-film signal was used as the positron trigger, and the energy deposited in each BGO channel was reconstructed from the pedestal-subtracted integrated charge. The energy calibration was performed using the characteristic gamma-ray features from the $^{22}$Na source, including the 511 keV annihilation peak and the 1275 keV prompt gamma-ray peak. The calibrated spectra were then compared channel by channel to check the uniformity of detector response.

\begin{figure}[t]

\centering

\includegraphics[width=0.95\textwidth]{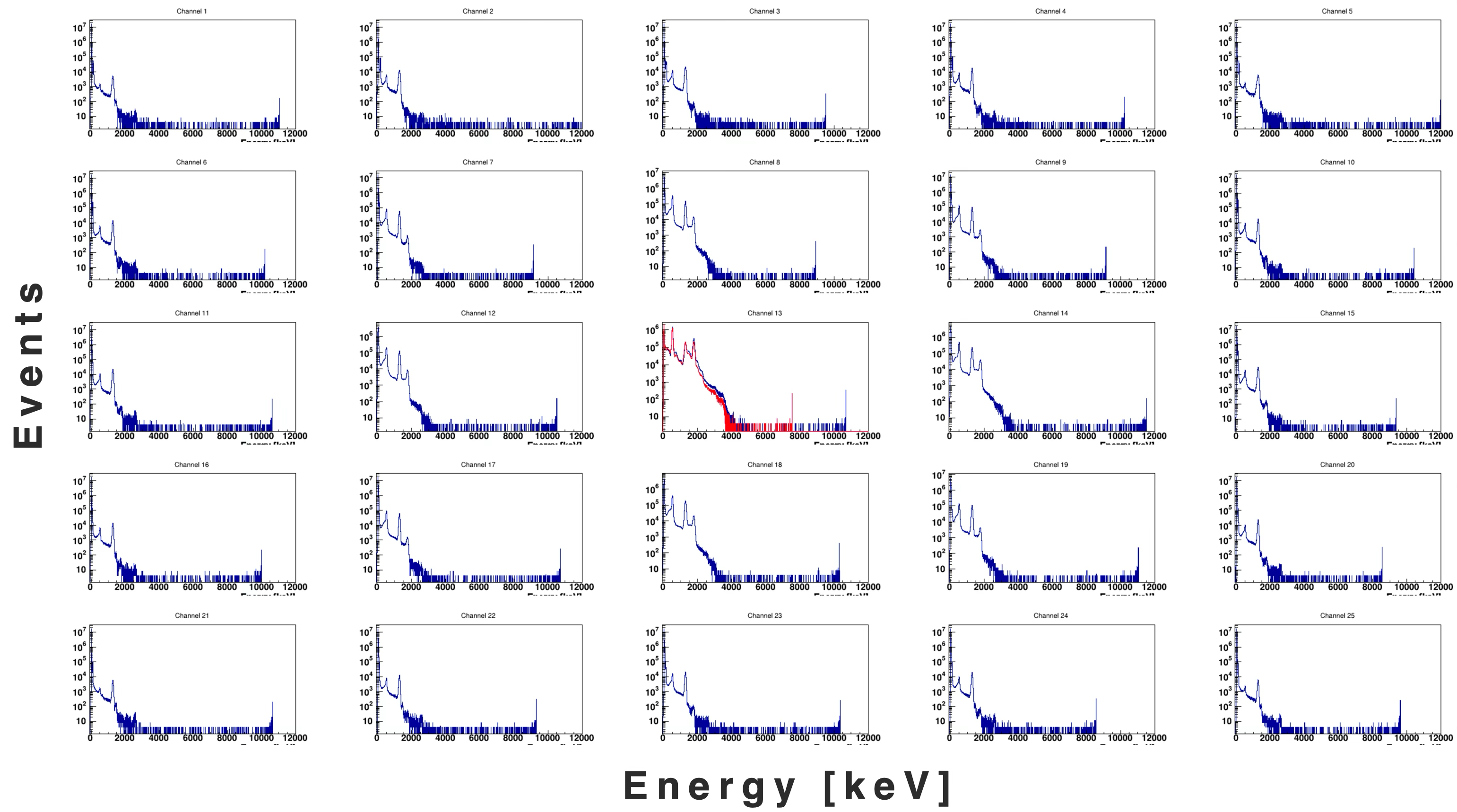}

\caption{Energy spectra measured at ground level over a one-hour period, shown for each channel of the $5 \times 5$ BGO array (channels 1--25). A high-energy continuum extending beyond the 1786~keV sum peak is visible in all channels and is attributed mainly to cosmic-ray-induced background events.}

\label{fig:ground_spectra}

\end{figure}

At ground level, the BGO spectra showed the expected gamma-ray structures from the $^{22}$Na source, as shown for all 25 channels in Figure~\ref{fig:ground_spectra}. In addition to the 511 keV and 1275 keV peaks, a sum peak around 1786 keV was observed when the 511 keV and 1275 keV gamma rays were deposited in the same channel. However, a high-energy continuum extending beyond the sum-peak region was also observed. This component is attributed mainly to cosmic-ray-induced events, which can deposit large energies in the BGO crystals and produce broad spectral structures. Even with the positron-gamma coincidence trigger applied, this background remained at the level of roughly 1 Hz, and occasional bursts of cosmic-ray activity raised the trigger rate to as much as about 65 Hz over a one-hour period.

The ground-level data illustrate both the basic functionality of the detector and the limitation of surface operation for an invisible new particle search. Although the detector response to the source was measured as expected, the high-energy continuum and environmental background increase the probability of abnormal event topologies. Since candidate events for the search are selected from rare missing-energy-like signatures, such backgrounds must be suppressed as much as possible. This motivated the underground implementation of the detector described in the previous section.

\subsection{Underground measurement}

After the ground-level commissioning, the detector was installed in the Yemi underground laboratory and operated inside the cryogenic shielding system. The underground measurement was performed under low-temperature and low-background conditions. The detector temperature was stabilized near $-37.5^{\circ}\mathrm{C}$, and the BGO signals were reconstructed using the charge integration and pedestal correction procedure optimized for the longer scintillation decay time at low temperature.

\begin{figure}[t]

\centering

\includegraphics[width=0.95\textwidth]{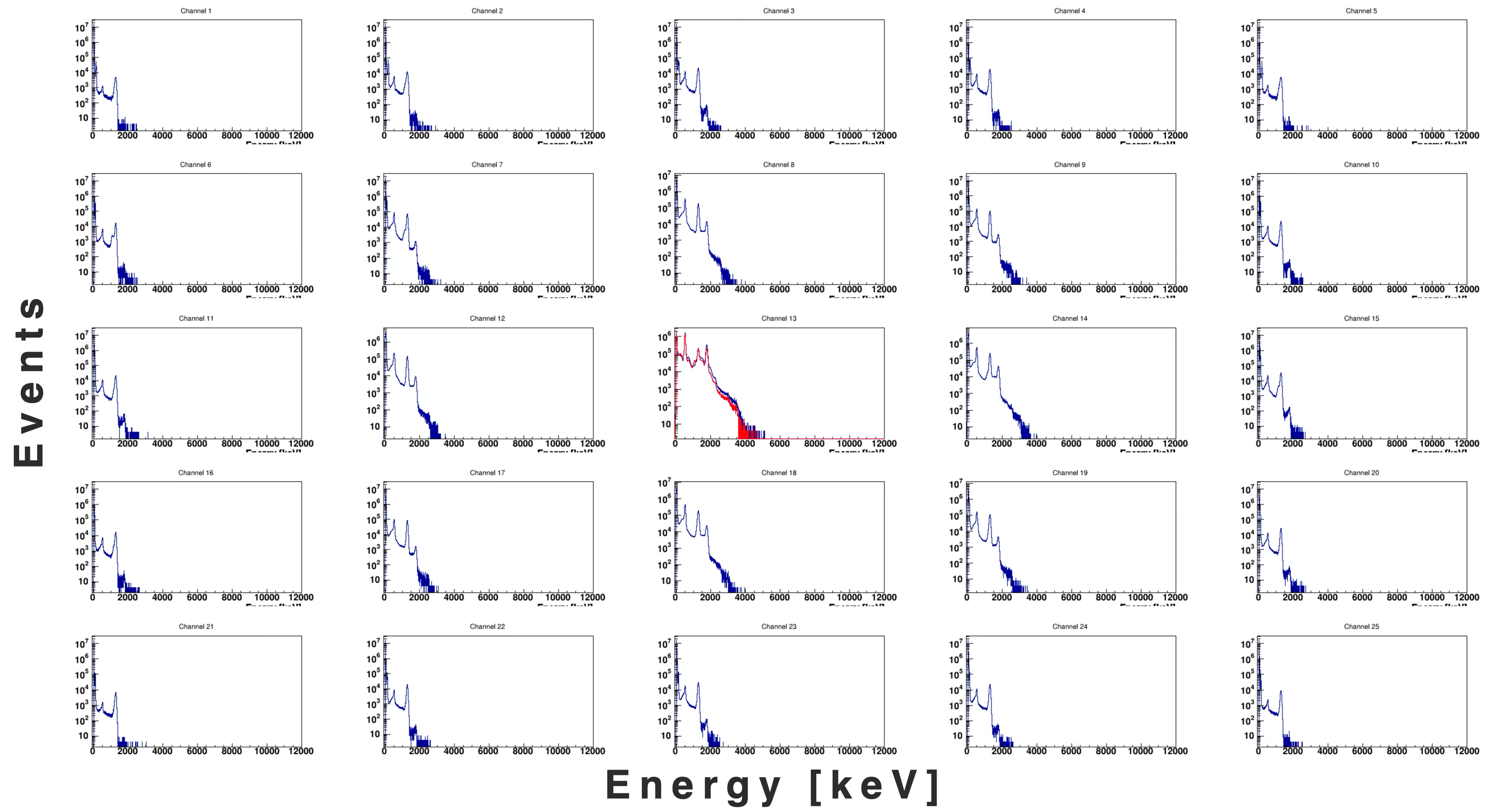}

\caption{Energy spectra measured underground over a one-hour period, shown for each channel of the $5 \times 5$ BGO array (channels 1--25). The high-energy continuum observed in the ground-level measurement (Figure~\ref{fig:ground_spectra}) is largely absent here.}

\label{fig:underground_spectra}

\end{figure}

The underground spectra, shown in Figure~\ref{fig:underground_spectra} for all 25 channels, show a marked reduction of the high-energy continuum observed at ground level. This indicates that cosmic-ray-induced background events are suppressed by the underground environment. Correspondingly, the coincidence-trigger background rate fell from the order of 1 Hz, with bursts up to about 65 Hz, at ground level to below approximately $3 \times 10^{-4}$~Hz underground, with no comparable bursts observed during the underground running period.

\begin{figure}[t]

\centering

\includegraphics[width=0.85\textwidth]{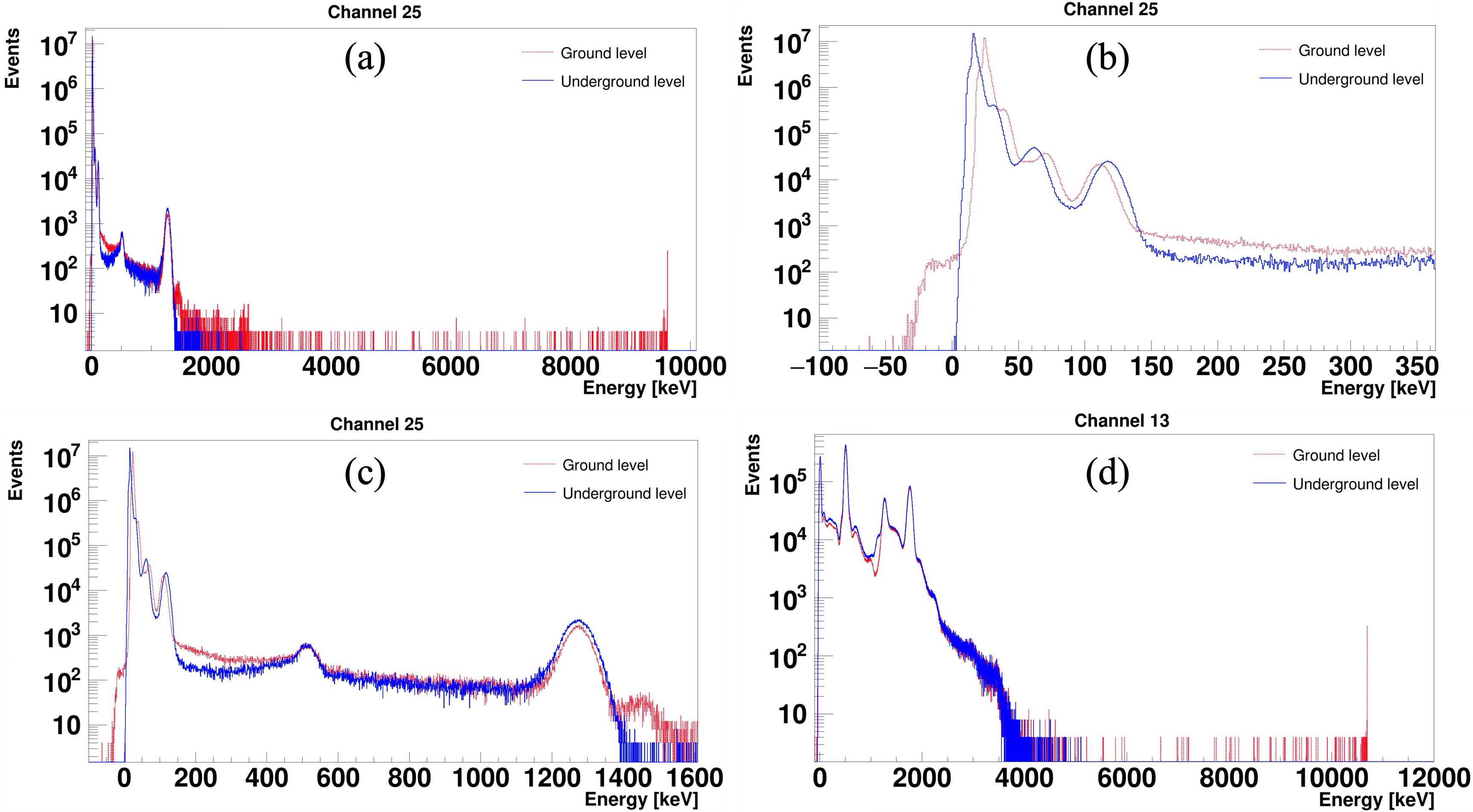}

\caption{Comparison of BGO spectra measured at ground level and underground. (a--c) Outermost channel (channel 25), with panel (b) showing a zoomed view of the low-energy region: the underground data show reduced low-energy background and improved noise stability relative to ground level. (d) Central endcap channel (channel 13), where a spectral shift near 1000~keV is attributed to an increase in SiPM gain at lower temperature.}

\label{fig:ground_underground_comparison}

\end{figure}

The underground measurement also improved the stability of the low-energy region. In the comparison between ground-level and underground data shown in Figure~\ref{fig:ground_underground_comparison}, representative BGO channels showed a reduction of low-energy noise and a more stable baseline. This improvement is attributed to the combined effect of underground operation, shielding, and reduced SiPM dark noise at low temperature. Low-energy stability matters here because the invisible new particle search is sensitive to small energy deposits and missing-energy-like event topologies.

Some temperature-dependent changes in the detector response were also observed. In particular, the central endcap BGO channel showed a spectral shift near the MeV region under low-temperature operation. This behavior is consistent with the temperature dependence of the SiPM gain and the BGO scintillation response. Therefore, channel-by-channel calibration and monitoring were required for the underground dataset. The observed shift does not prevent detector operation, but it must be accounted for in the comparison between experimental data and simulation.

The pedestal stability was evaluated using the pre-trigger waveform region, in the same way as described in Section~\ref{sec:Electronics} and shown in Figure~\ref{fig:pedestal}. Under the underground and low-temperature conditions, the pedestal values remained stable and well controlled, indicating that the real-time pedestal correction procedure worked as intended.

The underground measurement shows that the KAPAE Phase II detector operates as a stable cryogenic calorimeter in a low-background environment. Compared with ground-level operation, the underground setup reduced cosmic-ray-induced high-energy events and improved the low-energy baseline stability. Achieving part-per-billion (ppb) sensitivity for rare, invisible new particle searches requires an ultra-low-background environment to prevent false signals, and the KAPAE Phase II setup, installed deep underground with shielding, will provide promising experimental results for such searches.

\subsection{Energy resolution and comparison with previous experiments}

The energy resolution of the assembled KAPAE Phase II calorimeter was determined from the 511~keV and 1275~keV $^{22}$Na peaks, averaged over the 26 BGO channels, both at room temperature and after underground cryogenic operation. Table~\ref{tab:energy_resolution} summarizes the results. Cooling the detector to $-37.5^{\circ}\mathrm{C}$ improved the average FWHM energy resolution from $14.8 \pm 1.6\%$ to $11.7 \pm 1.5\%$ at 511~keV, and from $7.9 \pm 0.8\%$ to $6.5 \pm 0.5\%$ at 1275~keV, consistent with the increase in BGO light yield at low temperature discussed in Section~\ref{sec:BGO}.

\begin{table}[t]
\centering
\caption{Average BGO energy resolution (FWHM) at room temperature and after underground cryogenic operation ($-37.5^{\circ}\mathrm{C}$), for the 511~keV and 1275~keV $^{22}$Na peaks, averaged over the 26 detector channels.}
\begin{tabular}{lcc}
\hline
Energy [keV] & Room temperature [\%] & Underground, $-37.5^{\circ}\mathrm{C}$ [\%] \\
\hline
511  & $14.8 \pm 1.6$ & $11.7 \pm 1.5$ \\
1275 & $7.9 \pm 0.8$  & $6.5 \pm 0.5$ \\
\hline
\end{tabular}
\label{tab:energy_resolution}
\end{table}

This energy resolution is better than that of previous BGO-based invisible new particle searches. For individual diffused BGO crystals at 662~keV, the KAPAE Phase II crystals give an FWHM energy resolution of 10.8\%, compared with approximately 15\% reported for the crystals used in the ETH Zurich apparatus~\cite{gendotti}. This difference is attributed mainly to the chemically diffused crystal surface described in Section~\ref{sec:BGO}~\cite{diffuse}.

The overall detector performance reflects several design choices relative to earlier invisible new particle searches. The ETH Zurich experiment used an external plastic-scintillator positron trigger and operated its $4\pi$ BGO calorimeter at ground level~\cite{ethresult,crivelli2006}. The KAPAE Phase II detector instead uses an internal, SiPM-read-out PEN-film trigger placed directly next to the source, and operates underground with cryogenic cooling and lead/copper shielding. This combination gives both an improved energy resolution (Table~\ref{tab:energy_resolution}) and a coincidence-trigger background rate more than three orders of magnitude below that at ground level (Section~\ref{sec:Performance}). As discussed in Section~\ref{sec:Geant4}, these improvements are expected to give the detector the branching-ratio sensitivity needed to extend the current best limit for this class of search.

\subsection{Simulation-data comparison}

The detector response was further evaluated by comparing the underground data with Geant4 simulations. This comparison is necessary because the invisible new particle search depends on the accurate modeling of ordinary positronium and $^{22}$Na-related events, since events that produce low total energy in the BGO calorimeter can otherwise be mistaken for missing-energy signatures.

The Geant4 simulation included the detector geometry, BGO array, endcap crystals, source position, and trigger configuration. The simulated energy deposits were processed using detector-response corrections before being compared with experimental spectra. Channel-dependent energy resolution effects were applied to reproduce the measured peak widths, and the nonproportional scintillation response of BGO~\cite{intrinsicBGO} was also taken into account, since it can affect the reconstructed spectrum in the low-energy region.

\begin{figure}[t]

\centering

\includegraphics[width=0.85\textwidth]{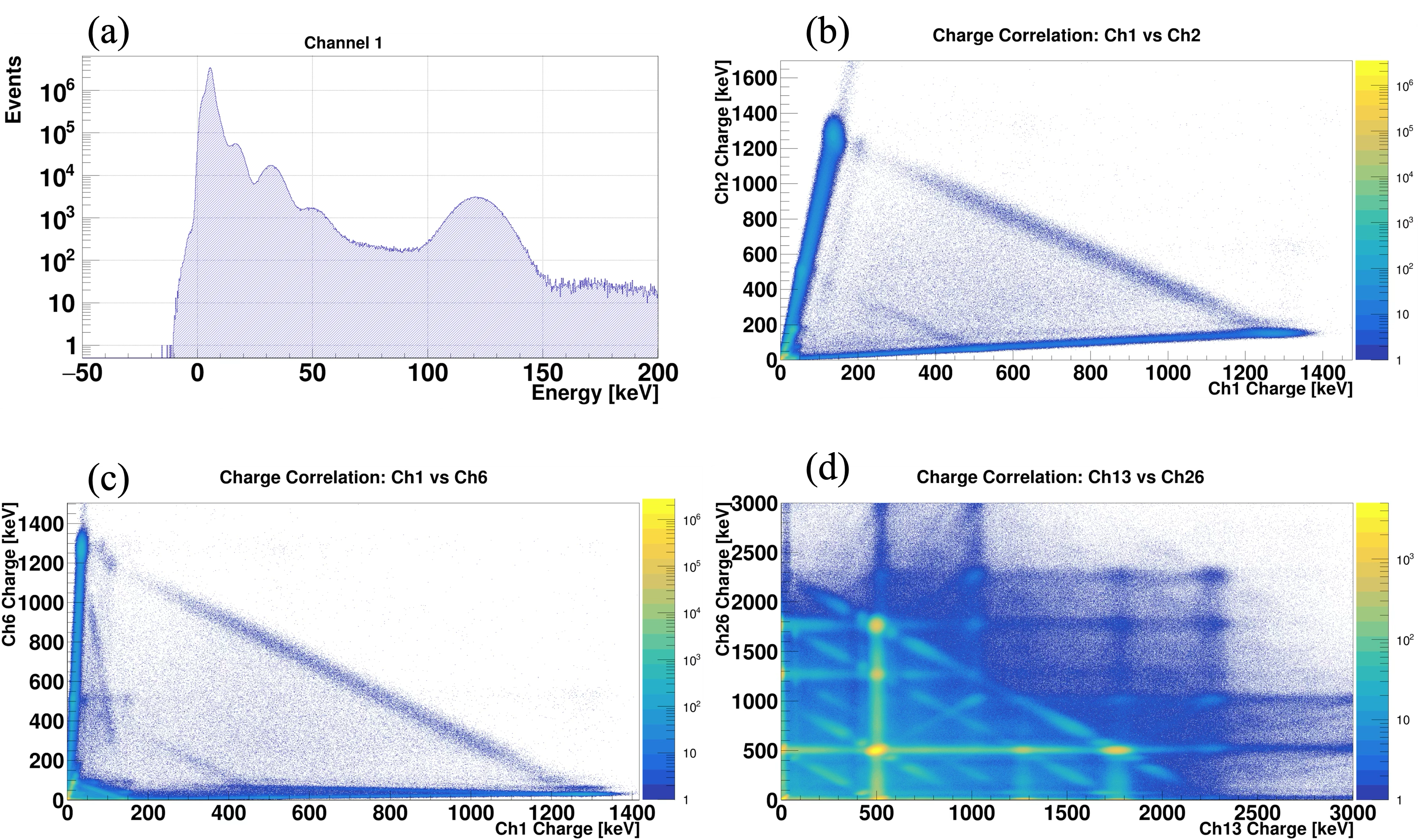}

\caption{Low-energy correlations between BGO channels. (a) Low-energy spec trum of channel 1, below 200~keV. (b, c) Charge correlations between channel 1 and its optically neighboring channels (channels 2 and 6), showing a diagonal correlation attributable to light leakage. (d) Charge correlation between the two endcap channels (13 and 26), which are not optically adjacent and show no comparable correlation.}

\label{fig:low_energy_correlation}

\end{figure}

As noted above, the underground operation suppressed the high-energy cosmic-ray background relative to ground level. In an early comparison with the simulation, however, an excess of low-energy events remained in the underground data. This excess was traced to optical light leakage between neighboring BGO crystals, examined through the channel-to-channel correlations shown in Figure~\ref{fig:low_energy_correlation}. Low-energy correlations between optically adjacent crystals indicate that a fraction of scintillation light leaks from one channel to another, producing small apparent energy deposits in neighboring channels; the correlation between the two endcap channels, which are not optically adjacent, shows no such feature. These correlations provided the basis for a light-leakage correction applied in the event reconstruction.

\begin{figure}[t]

\centering

\includegraphics[width=0.95\textwidth]{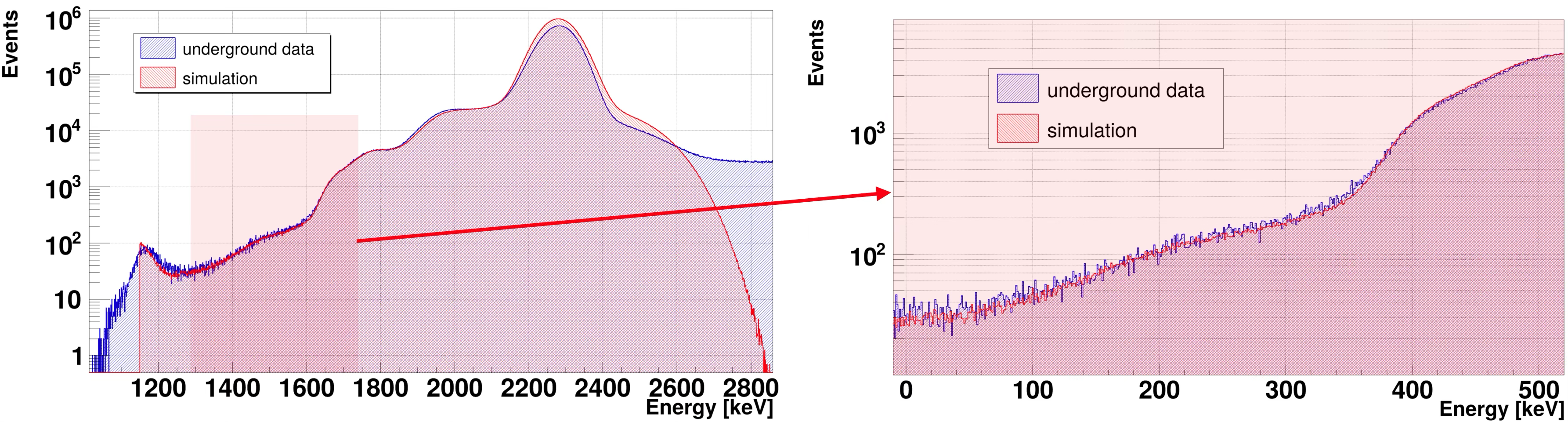}

\caption{Total energy deposited across the full BGO array, summed over all channels, for underground data (blue) and Geant4 simulation (red). Left: overall comparison. Right: expanded view of a lower-energy region of particular interest for the invisible new particle search.}

\label{fig:total_energy_comparison}

\end{figure}

After this correction, the total energy deposited across the full BGO array was compared with the Geant4 simulation, as shown in Figure~\ref{fig:total_energy_comparison}, together with an expanded view of a lower-energy region of particular interest for this search. The total energy deposit agrees well with the simulation in both panels, supporting the use of the simulation for estimating ordinary-event backgrounds in the invisible new particle search.

\section{Conclusions}
\label{sec:Conclusions}

The KAPAE Phase II detector was developed as a compact calorimetric system for an invisible new particle search in total and partial invisible decays of positronium. The final detector consists of a $5 \times 5$ BGO array, mostly of dimensions $30 \times 30 \times 150~\mathrm{mm}^{3}$, with two shortened endcap crystals accommodating a central $^{22}$Na source and PEN-film positron trigger. Compared with the segmented Phase I detector, this compact, reduced-channel geometry lowers the probability that ordinary annihilation events are misreconstructed as missing-energy-like background. Geant4 simulations guided both the calorimeter geometry and the PEN-film trigger thickness, which was set to three $125~\mu\mathrm{m}$ layers as a compromise between positron-tagging efficiency and gamma-ray attenuation.

The BGO crystals, with a chemically diffused surface, together with the custom $4 \times 4$ SiPM array, preamplifier, and DAQ system, form an integrated, low-noise readout chain with event-by-event pedestal correction. The measured energy resolution is 10.8\% FWHM at 662~keV for a single crystal, and 11.7\%/6.5\% at 511/1275~keV averaged over the assembled detector at $-37.5^{\circ}\mathrm{C}$; this improves on the $\sim$15\% typical of earlier BGO-based invisible new particle searches~\cite{gendotti} (Section~\ref{sec:Performance}). Together with the reduced-segmentation geometry and the underground, shielded, cryogenic operation, this resolution is expected to give KAPAE Phase II the branching-ratio sensitivity projected in Section~\ref{sec:Geant4}.

The assembled detector was installed in the 1000~m deep Yemi underground laboratory, inside a lead- and copper-shielded cryogenic chamber stabilized at $-37.5 \pm 0.5^{\circ}\mathrm{C}$. This suppressed the positron-gamma coincidence background from order 1~Hz at ground level, with bursts up to about 65~Hz, to below $3 \times 10^{-4}$~Hz underground. After correcting for optical light leakage between neighboring BGO channels, the total underground energy spectrum agrees well with the Geant4 simulation, supporting its use for background estimation in the invisible new particle search.

In summary, the KAPAE Phase II detector was optimized and installed as intended for an invisible new particle search in positronium decay, meeting the design requirements for geometry, energy resolution, and background level. The corresponding invisible new particle search will be reported separately.




\end{document}